\documentstyle[amssymb,preprint,aps,epsf,graphicx,psfig]{revtex}

\newcommand{\bff}[1]{{\mbox{\boldmath $#1$}}}

\begin{document}
\title{Exploration of  Resonant Continuum and Giant Resonance in the
Relativistic Approach}
\author{Li-gang Cao  and Zhong-yu Ma\thanks{also Center of Nuclear Theoretical Physics,
National Laboratory of Heavy Ion Accelerator, Lanzhou 730000,
China and Institute of Theoretical Physics, Academia Sinica,
Beijing 100080, China}}
\address{  China Institute of Atomic Energy, Beijing 102413, China }
\maketitle

\begin{abstract}
Single-particle resonant-states in the continuum are determined by
solving scattering states of the Dirac equation with proper
asymptotic conditions in the relativistic mean field theory (RMF).
The regular and irregular solutions of the Dirac equation at a
large radius where the nuclear potentials vanish are relativistic
Coulomb wave functions, which are calculated numerically.
Energies, widths and wave functions of single-particle resonance
states in the continuum for $^{120}$Sn are studied in the RMF with
the parameter set of NL3. The isoscalar giant octupole resonance
of $^{120}$Sn is investigated in a fully consistent relativistic
random phase approximation. Comparing the results with including
full continuum states and only those single-particle resonances we
find that the contributions from those resonant-states dominate in
the nuclear giant resonant processes.

\end{abstract}


\section{Introduction}
In recent years, the physics of exotic nuclei attracts more
attentions both experimentally and theoretically. Due to the
closeness of the Fermi surface to the particle continuum in exotic
nuclei  the coupling between bound states and the particle
continuum become important. The description of exotic nuclei has
to explicitly include the coupling between bound states and the
continuum\cite{Ring96,Dob98}. Although there are several extensive
theoretical studies of exotic nuclei using relativistic and
non-relativistic microscopic methods\cite{Ring96,Dob98,Kuo97}, a
rigorous treatment of the continuum for exotic nuclei is still
under investigation. Nuclei beyond the drip line are usually
unbound and often exhibit resonances with a pronounced
single-particle character, which are single-particle resonance
states. The direct emission of charged particle from an unbound
ground state has been observed in the three-body decay of
$^{12}$O\cite{Kry95}, which is closely linked with the energies
and widths of resonant-states. In microscopic studies nuclear
collective giant resonances are explained by the particle-hole
excitations in a coherent way, where the particle continuum plays
an important role\cite{Ma97,Ma01,Ring01}. The contribution from
the resonant states in the giant resonances has been studied in
the so-called resonant random phase approximation in the
non-relativistic approach by Curutchet et. al.\cite{Cur89}. It was
found that the contribution of the particle continuum is mainly
from the single-particle resonances in the continuum. Therefore,
the exploration of single-particle resonant-states becomes very
important in explaining the properties of ground states, dynamical
processes of nuclei, excited nuclei, especially in understanding
properties of nuclei far from the $\beta$ stable line.

It is known that in most previous theoretical nuclear structure
calculations, when the contribution from the continuum is
performed the single-particle states in the continuum are usually
treated by a discretization procedure with expanding the wave
functions in a harmonic oscillator basis or setting a
box\cite{Ring96,Ma01}. This approximation, however, can be
justified for very narrow resonances and gives a global
description of the contributions from the continuum. It is
desirable to explore the resonant continuum and look for efficient
methods for the description of single-particle resonance states.
There are several techniques developed to explore resonant-states
in the continuum. A possible approach is the reaction method based
on R-matrix theory. Analyzing the R-matrix and using the relation
to the collision matrix permit one to define the energies and
widths of resonant-states\cite{Var92}. This method is mainly used
to analyze experimental data or within phenomenological models for
simple interactions and systems. For composite many-body systems,
the real stabilization method\cite{Haz70} by solving the
Schr\"{o}dinger equation in changing the box's size within certain
limits is implemented for finding the energies and widths of
resonant-continuum states. The complex scaling method combined
with the Hartree-Fock method\cite{Kru97} is thought as an
effective method in exploring wider resonant-states. It is
performed by rotating the coordinate $r$ by an unbounded
non-unitary scaling operator and transforming the continuum
resonance wave function to the normalizable wave function of the
bound state. Recently, the authors of Ref.\cite{Tan97} apply the
analytic continuation in the coupling constant method to find
energies and widths of resonant-states. Most of works described
above are performed only in the framework of non-relativistic
approach. Recently relativistic mean field models (RMF) have been
successfully applied to a variety of problems in nuclear structure
and reactions.  To deal with the continuum one requires the
knowledge of the resonant in the continuum in the relativistic
approach. It is the main purpose of this paper to explore the
single-particle resonant-states in the framework of RMF models.

In the present work, we aim at the investigation of the
resonant-states by S-matrix method in the RMF. The scattering
states of Dirac equation are treated with a proper asymptotic
behavior. The regular and irregular relativistic Coulomb wave
functions for the Dirac equation can be derived analytically.
Energies, widths and wave functions of single particle resonances
in the continuum are solved consistently in the coordinate space.
The energy of a resonant-state is determined when the phase shift
of the scattering state reaches $\pi /2$. The width of a
resonant-state is the full width of half magnitude (FWHM) for the
corresponding partial wave scattering cross section. The S-matrix
method is a straightforward and feasible method, which is suitable
for various cases, light and heavy, stable and exotic nuclei. We
shall study the single-particle resonant-states for $^{120}$Sn in
this paper. In order to verify the importance of exploring the
resonances in the continuum we study the response function of
isoscalar octupole response for $^{120}$Sn in the framework of a
fully consistent relativistic random phase approximation based on
the RMF. By analyzing the distribution of the octupole strength,
we will show that the contributions in the giant resonance
processes of nuclei mainly come from the single-particle
resonant-states in the continuum.
Our results are consistent with the previous studies of giant
resonances by means of the so-called resonant random phase
approximation in the non-relativistic approach by Curutchet et.
al.\cite{Cur89}, where the single particle basis is composed of
single particle bound and resonant states.

The paper is arranged as the follows. In Sec.II the S-matrix
methods in the Dirac equation is presented. The Single particle
resonances of $^{120}$Sn are given in Sec.III. We study the
Octupole resonance of $^{120}$Sn with those single particle
resonances in the continuum in Sec. IV. Finally we give a brief
summary in Sec. V.

\section{S-matrix method in the RMF}

The Dirac equation of a single-particle state in the RMF
approximation can be expressed as following,
\begin{equation}
\left[ {\bff{\alpha}} \cdot {\bff{p}}+V(r)+\beta \left(
M-S(r)\right) \right] \psi _i=E_i\psi _i~,  \label{eq1}
\end{equation}
where  $E_i$ and $\psi _i$ are the single-particle energy and wave
function, respectively. $S(r)$ and $V(r)$ are an attractive scalar
and repulsive vector potential, respectively. In the RMF
approximation, scalar and vector potentials are produced by the
classical meson fields: isoscalar $\sigma$, $\omega$ and isovector
$\rho$ mesons as well as the photon, which are obtained in a
self-consistent calculation for the nuclear ground
state\cite{Ring96}. The nucleon spinor $\psi_i$ in the Dirac
equation is expressed as,
\begin{equation}
\psi _\alpha ({\bff{r}})=\frac{1}{r} \left( \begin{array}{c}
  ig_a(r) \\
  f_a(r){\bff{\sigma}}\cdot {\bff{\hat{r}}}
\end{array} \right)
\Phi_{\kappa,m} ( \hat{r})\chi _{\frac 12}~, \label{eq2}
\end{equation}
where  $\alpha $ is a set of quantum numbers $\alpha
=(n,\ell,j,m,\tau _3)\equiv(a,m,\tau_3)$,
$\kappa=(-1)^{j-\ell+1/2}(j+1/2)$, $\chi _{\frac 12}$ is the
isospinor. For spherical nuclei, the Dirac equation can be reduced
to coupled equations of the radial part for $g_a(r)$ and $f_a(r)$.
\begin{equation}
\frac{d}{dr}
  \left( \begin{array}{c}
    g_a(r) \\
    f_a(r) \
  \end{array} \right) =\left(\begin{array}{cc}
    -\frac{\kappa}{r} & M+E_a-S(r)-V(r)) \\
    M-E_a-S(r)+V(r) & \frac{\kappa}{r} \
  \end{array}\right)\left(\begin{array}{c}
    g_a(r) \\
    f_a(r) \
  \end{array}\right)~.
\label{eq3}
\end{equation}
Single-particle bound states as well as the scalar and vector
potentials for the nuclear ground state are determined
self-consistently by solving the Dirac equation in an iterative
way. With those scalar and vector potentials unoccupied bound
states can be calculated straightforward. For scattering states, a
proper asymptotic condition for each partial wave has to be
imposed. At a large distance, say $R$ = 15 fm for light nuclei and
20 fm for heavy nuclei, nuclear potentials generated by exchanging
mesons vanish, and only Coulomb potential remains. In the
relativistic approach a spin dependent interaction even with only
Coulomb interaction is built into the theory automatically.
Therefore Coulomb functions of the Dirac equation are different
from those in the non-relativistic approach. The relativistic
treatment of the Coulomb functions are rather difficult task,
which is described in detail in Refs.\cite{Gre90,Ros61}. Two
independent solutions of real Coulomb radial functions are denoted
by $G^r_\kappa(r)$, $F^r_\kappa(r)$ corresponding to $g_a$ and
$f_a$ in the Eq.\ref{eq2} for the solution regular at the origin
and by $G^{ir}_\kappa(r)$, $F^{ir}_\kappa(r)$ for the solution
irregular at the origin.
\begin{eqnarray}
G^r_\kappa(r)&=&\frac{ \sqrt{E+M}(2kr)^{\gamma}e^{-\pi \eta
/2}|\Gamma(\gamma-i\eta)|}{2\sqrt{\pi k}\Gamma(2\gamma+1)} \{
e^{i(-kr+\phi)}(\gamma-i\eta)F(\gamma+1-i\eta,2\gamma+1;2ikr)+c.c.
\}~, \label{eq4}\\
 F^r_\kappa(r)&=&\frac{i \sqrt{E-M}(2kr)^{\gamma}e^{-\pi
\eta /2}|\Gamma(\gamma-i\eta)|}{2\sqrt{\pi k}\Gamma(2\gamma+1)} \{
e^{i(-kr+\phi)}(\gamma-i\eta)F(\gamma+1-i\eta,2\gamma+1;2ikr)-c.c.\}~,\nonumber
\end{eqnarray}
where  $\gamma^2=\kappa^2-Z^2\alpha^2$, $\eta=Z\alpha E/k$ and the
phase $\phi$ is determined by $e^{2i\phi}=-\frac{\kappa+i\eta
M/E}{\gamma-i\eta}$. $\alpha$ is the fine structure constant and
$k$ is the wave number $k^2=E^2-M^2$.
$F(\gamma+1-i\eta,2\gamma+1;2ikr)$ is the confluent hypergeometric
function. $c.c.$ denotes the complex conjugate. The irregular
solutions are obtained from those in Eq.\ref{eq4} by the change of
the sign of $\gamma$. The asymptotic behavior of regular and
irregular solutions are
\begin{eqnarray}
G_\kappa ^r(r) &\longrightarrow& \sqrt{\frac{k}{\pi(E-M)}}
\cos(kr+\delta_\kappa)~,~~~~~ F_\kappa ^r(r) \longrightarrow
-\sqrt{\frac{E-M}{\pi k}} \sin(kr+\delta_\kappa)~, \label{eq5}\\
G_\kappa ^{ir}(r)&\longrightarrow& -\sqrt{\frac{k}{\pi(E-M)}}
\sin(kr+\delta_\kappa)~,~~~~~ F_\kappa ^{ir}(r)\longrightarrow
-\sqrt{\frac{E-M}{\pi k}} \cos(kr+\delta_\kappa)~. \\ \nonumber
\end{eqnarray}
The Coulomb phase shift $\delta_\kappa$ has a slightly different
expression as that in the non-relativistic case,
$\delta_\kappa=\phi-\eta \ln 2kr -\frac \pi 2 \gamma- \arg
\Gamma(\gamma-i \eta)$, which depends not only on the orbital
angular momentum $\ell$ but also on the total angular momentum
$j$. It is found that deviations of the upper components in the
Dirac Coulomb functions from the Coulomb wave functions in the
non-relativistic approach are pronounced in the case that the
nucleus is very heavy (for example in the extreme case $Z \sim
1/\alpha$ and $|\kappa| = 1$). For the neutron case, they reduce
to spherical Bessel functions.

In practice, the scattering solution of the Dirac equation regular
at the origin is obtained by integrating the coupled equations of
radial part from the origin to a certain distance $R$, where the
nuclear potential vanishes. The boundary condition is imposed and
the upper component of the Dirac radial wave function has the
following proper asymptotic behavior:
\begin{equation}
g_a(kr)=A_a [ G_\kappa ^r (kr)-\tan \Delta _a  G_\kappa
^{ir}(kr)]~,  ~~~~~ {\textrm{for}}~~  r\geq R~. \label{eq7}
\end{equation}
The phase shift $\Delta _a$ of the corresponding partial wave can
be uniquely determined by  matching the inner wave function.

The resonant-state is characterized by the phase shift $\Delta _a$
crossing to $\pi /2$, where the scattering cross section of the
corresponding partial wave reaches its maximum. Those
single-particle resonant-states are meta-stable states captured by
Coulomb(for proton) and centrifugal barriers, which wave functions
have large probabilities inside the nuclear potential. For the
proton, it is easy to form resonant-states due to the presence of
an appreciably high Coulomb barrier outside of the nucleus. In
contrast, for the neutron the resonant-state with a low orbital
angular momentum does not always exist because of the lack of a
sufficiently high centrifugal barrier. The decay widths of those
resonant-states can be roughly explained by considering the
penetrability through the Coulomb and centrifugal barriers in view
of quantum mechanics. The width of the resonant-state may be
determined by the FWHM of the corresponding partial wave
scattering cross section.

\section{Single particle resonances in the continuum for
$^{120}$Sn}

We investigate the single particle resonant-states in the
continuum for $^{120}$Sn in terms of  the S-matrix method in the
RMF approximation described above. The single-particle energy
levels for $^{120}$Sn are shown in Fig.1 together with the central
potentials. All results are calculated with NL3 parameter
set\cite{Lal97}. Those states with positive energy in the
continuum in Fig.1 are the resonant-states we have found.
Energies and their widths of those single particle resonances for
$^{120}$Sn are listed in Table 1.
It is found that some states with very small positive energies,
such as neutron 1h$_{9/2}$ state (0.229+$i 4.4 \times 10^{-8}$)
MeV are very narrow. It is due to the presence of appreciably high
and thick centrifugal (and Coulomb) barriers. Those states are
rather stable and have bound state characteristics.

The effect of the Dirac Coulomb wave function for $^{120}$Sn with
Z=50 is investigated. The relativistic effects on the Coulomb wave
functions are not large due to the fact that the nuclide Sn is not
too heavy and $Z \ll 1/\alpha$. It is found that the largest
deviation of the resonance energy for the $\kappa=1$ state is
within a few percent if we use non-relativistic Coulomb wave
functions for the upper component of the Dirac spinor instead.

In addition to those resonant-states, some state near the top of
the barrier has the following properties: the partial wave cross
section has a resonant character, which reaches its maximum value,
while the phase shift closes to, but not reaches $\frac \pi 2$.
Those states may be called as resonant-like states, such as proton
state 2g$_{7/2}$ for $^{120}$Sn.

\section{Octupole resonance of $^{120}$Sn in the RRPA}

 Recently, a fully self-consistent relativistic random
phase approximation (RRPA) \cite{Ma01,Ring01} has been established
based on the RMF approximation and has achieved a success in
describing the multipole giant resonance properties of closed
shell nuclei\cite{Ma02}. In those investigations, the particle
continuum is treated by a discretization procedure with expanding
wave functions in a harmonic oscillator basis. An energy average
with a Lorentzian factor was assumed by hand.  It is known from
the studies in non-relativistic approaches\cite{Cur89} that
contributions of the continuum to the giant resonances are mainly
from those single-particle resonances. In this paper we perform
the RRPA calculation with the continuum replaced by a set of
single-particle resonances. The wave functions of the
resonant-states are localized by solving the Dirac equation in the
coordinate space with a proper box, which position is determined
by reproducing the energy of each single-particle resonance. An
advantage of this calculation is that single particle escaping
widths in the giant resonance are obtained automatically. It
allows one to obtain the desired quantities very fast. One can
even gain some physical insight on those quantities. Of course it
is known that a complete treatment of the continuum may be carried
out by the continuum RRPA with a Green function
method\cite{Weh93}.

The response function of a quantum system to an external field is
given by the imaginary part of the polarization
operator\cite{Ma97,MTG97},
\begin{equation}
R(Q,Q;{\bf k},{\bf k^{\prime }};E)=\frac 1\pi Im\Pi ^R(Q,Q;{\bf k},{\bf %
k^{\prime }};E)~,  \label{eq11}
\end{equation}
where Q is an external field operator. In the view of the
theoretical method, a giant resonance is the coherent
superposition of one particle-one hole states with definite spin
and parity.

The isoscalar giant octupole ($Q=r^3Y_{30}$) response (ISGOR) of
$^{120}$Sn has not been investigated in the RRPA, although the
experiment data are available\cite{Yam81}. In this paper we take
the ISGOR of $^{120}$Sn as an example to examine the contribution
of resonant-states in nuclear dynamical processes. We know that
the octupole excitation can be identified as two parts: $1\hbar
\omega $ and $3\hbar \omega $ excitations, which are called the
low-energy octupole resonance (LEOR) and high-energy octupole
resonance (HEOR), respectively. The LEOR contains a number of
configurations, which are mainly contributed from those single
particle resonances in the continuum. In the calculation the
particle continuum is replaced by the resonant-states only which
are described above. It has been pointed out \cite{Ma01,Ring01}
that the contribution of the negative energy Dirac states in the
RRPA can not be neglected in the isoscalar modes where the scalar
meson plays an important role. Therefore in the RRPA calculation
based on the RMF ground state the unoccupied states include not
only the unoccupied positive energy states, but also the negative
energy states in the Dirac sea.

In Fig.2, we show the response function of the ISGOR for
$^{120}$Sn calculated by including the single-particle
resonant-states with their widths in the continuum, which are
displayed by dashed curves. Response functions, where all
discretized states in the continuum with a Lorentzian factor
$\triangle =1$ MeV  are included\cite{Ma01}, are plotted with
solid curves.  The unperturbed and RRPA response strengths in two
methods are shown in Fig.2 (a) and (b), respectively. The
experimental data are plotted by arrows. A good agreement with
experiment data is found, except for the HEOR, where the
theoretical calculations predict slightly higher energy.

The Hartree strength in Fig.2(a) are the strength of incoherent
particle-hole excitations. The width of the dashed curve is rather
narrow  due to the widths of particle resonances, while the width
of the solid curve is produced by the energy average.
We analyze the contribution to the Hartree strength from various
particle-hole pairs. It is found that those pairs formed between
unoccupied and occupied bound states give the important
contribution ($\sim 54\%$) to the integral of the strength with
respect to the energy ($M_0=\int_0^{E_{max}} R(E) dE$) in the
unperturbed strength. Besides the contributions from the
excitations between bound states, the mostly rest strength is
produced from the excitations between the occupied hole states and
the positive energy single-particle resonant-states. Obviously
neutron particle-hole configurations are mostly substantial ($\sim
67\%$ in $M_0$), especially for the LEOR due to heavy nuclei with
more neutrons than protons. Among them the unoccupied states, e.g.
$\nu (2f_{7/2})$, $\nu (1h_{11/2})$ and single particle resonant
state $\nu (1h_{9/2})$, $\nu (2f_{5/2})$ with narrow widths
characterize the largest contributions. The contribution to $M_0$
from wide single particle resonant states , e.g. $\pi(2g_{9/2})$
are small, but not negligible. The low-lying strength at
$E_x\leqslant 7.5MeV$, is completely formed from the neutron
excitation between bound levels: $\nu(1h_{11/2}2d_{5/2}^{-1}) $
(3.083MeV), $\nu(1h_{11/2}1g_{7/2}^{-1}) $ (4.264MeV), $\nu (
2f_{7/2}3s_{1/2}^{-1})$ (6.957MeV), $\nu(2f_{7/2}2d_{3/2}^{-1})$
(7.328MeV), where the value in the parenthesis is its Hartree
energy. Naturally, strengths of those low-lying states are
discrete and have no width. The width shown in the figure is
obtained by expanding with a Lorentzian factor $\triangle =1$ MeV
in this calculation. It should be mentioned that the pairs formed
from the states in the Fermi sea and Dirac sea has negligible
contribution to the Hartree strength, but play an important role
in the coherent strength, which is discussed in more detail in
Refs.\cite{Ma01,Ring01}.

 In Fig.2(b) we show the correlated strength of the ISGOR for $^{120}$Sn.
The strengths are pushed down by the residual particle-hole
interaction through the meson exchanges, which clearly show a
collectivity in the RRPA strength. The LEOR with $1\hbar \omega$
excitations for $^{120}$Sn is located at $6\leqslant E_x\leqslant
20MeV$ and the HEOR with $3 \hbar \omega$ excitations is
identified as peaks around 27-38 MeV. Since we are working on the
nonspectral RPA representation, it is difficult to analyze the
detail contribution to the RPA strength from various
configurations. In order to make a clear analysis it is necessary
to reformulate and work in the configuration space, which are our
future works. In comparison with the strengths in Fig.2(a) it is
interesting to notice that some strengths, for example the
strength at the energy around 21 MeV, are missing in the dashed
curve, which are not produced from resonant-states,
$\nu(3h_{9/2}2d_{3/2}^{-1})$ and $\nu(3h_{11/2}2d_{5/2}^{-1})$.
Those strengths are not changed in the RRPA strength, see in Fig.
2(b). It is clearly shown that those states do not contribute to
the coherent interaction and the strength remains at the same
position as the Hartree strength.

For the coherent RRPA strengths both methods give similar results.
The one in the simple method including only resonant-states in the
continuum produces main features of giant resonances at the energy
below 40 MeV. It characterizes not only the Landau width, but also
the single particle escaping width, which may be rigorously
described by the continuum RRPA\cite{Weh93}. The results
illustrate that the contribution from the continuum to the
coherent strength in giant resonances is mainly from those single
particle resonances.

\section{A brief summary}

In a summary, we have studied energies, widths and wave functions
of single particle resonant-states in the continuum for $^{120}$Sn
using the S-matrix method based on the RMF with the parameter set
NL3. The scattering Dirac wave functions are calculated with a
proper asymptotic scattering boundary condition. The relativistic
Coulomb wave functions are investigated, which are different from
the non-relativistic ones, but the deviations are small for the
light and even intermediate heavy nuclei. The single-particle
resonance state in the continuum is defined when its scattering
phase shift passes through $\pi /2$ in the S-matrix. It shows that
the S-matrix method is a straightforward and feasible way in the
determination of single-particle resonances. The single particle
resonances in the continuum play an important role in the
description of the nuclear dynamical processes, such as the
collective giant resonances. In the framework of fully consistent
RRPA,  the ISGOR of $^{120}$Sn can be well described by the simple
method where the continuum replaced by only single particle
resonances. With this simple method we can not only obtain the
desired quantities very fast, but also gain some physical insight.

\vspace{1cm} {\bf ACKNOWLEDGMENTS}

The authors gratefully acknowledge stimulating discussions with
Prof. Xizhen Zhang. This work is supported by the National Natural
Science Foundation of China (10075080,19835010) and Major State
Basic Research Development Program under contract No. G20000774.

\newpage
\begin{table}
\caption{ Energies and widths of single particle resonant-states
in $^{120}$Sn. The results are calculated with NL3 parameter set.
All energy values in the table are in unit of MeV.}
\begin{tabular}{ccc}
 State & $E _p$ & $E_n$   \\
\hline
 2$f_{7/2}$ & 6.210+$i 0.043$ & \\
 1$h_{9/2}$ & 7.132+$i 0.003$ & 0.229+$i 0.000$ \\
 3$p_{3/2}$ & 7.513+$i 0.924$ &  \\
 2$f_{5/2}$ & 7.934+$i 0.307$ & 0.657+$i 0.031$ \\
 3$p_{1/2}$ & 8.085+$i 1.344$ & 0.176+$i 0.316$ \\
 1$i_{13/2}$ & 10.110+$i 0.012$ & 3.261+$i 0.004$ \\
 2$g_{9/2}$ & 14.772+$i 3.502$ & \\
 1$i_{11/2}$ & 16.960+$i 0.999$ & 9.751+$i 1.384$ \\
 1$j_{15/2}$ & &  12.658+$i 1.051$ \\
\end{tabular}
\end{table}

 \begin{figure}[tbp]
\includegraphics*[scale=0.5,angle=0.]{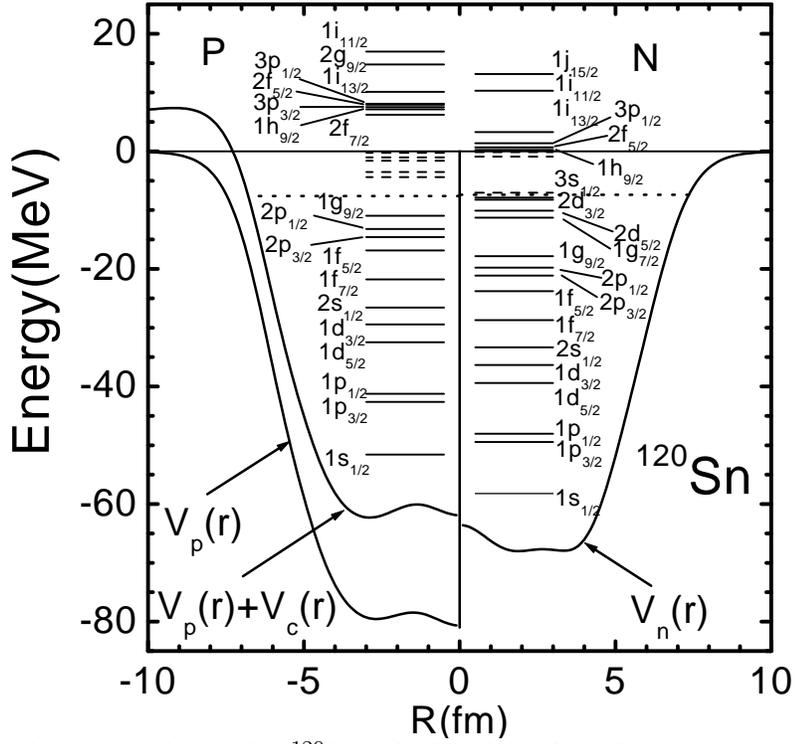}
\caption{Single-particle levels for $^{120}$Sn calculated in the
relativistic mean field theory. Those for protons (neutron) are
shown on the left (right) hand side. The dashed lines represent
the unoccupied bound levels, they are from the Fermi surface:
$1g_{\frac {7}2}$, $2d_{\frac {5}2}$,  $2d_{\frac {3}2}$,
$3s_{\frac {1}2}$, $1h_{\frac {11}2}$  for protons and $1h_{\frac
{11}2}$, $2f_{\frac {7}2}$, $3p_{\frac {3}2}$  for neutrons,
respectively. The dotted lines indicate the approximate positions
of the Fermi surfaces for the proton and neutron, respectively. }
\label{fig1}
\end{figure}

\begin{figure}
\includegraphics*[scale=0.6,angle=0.]{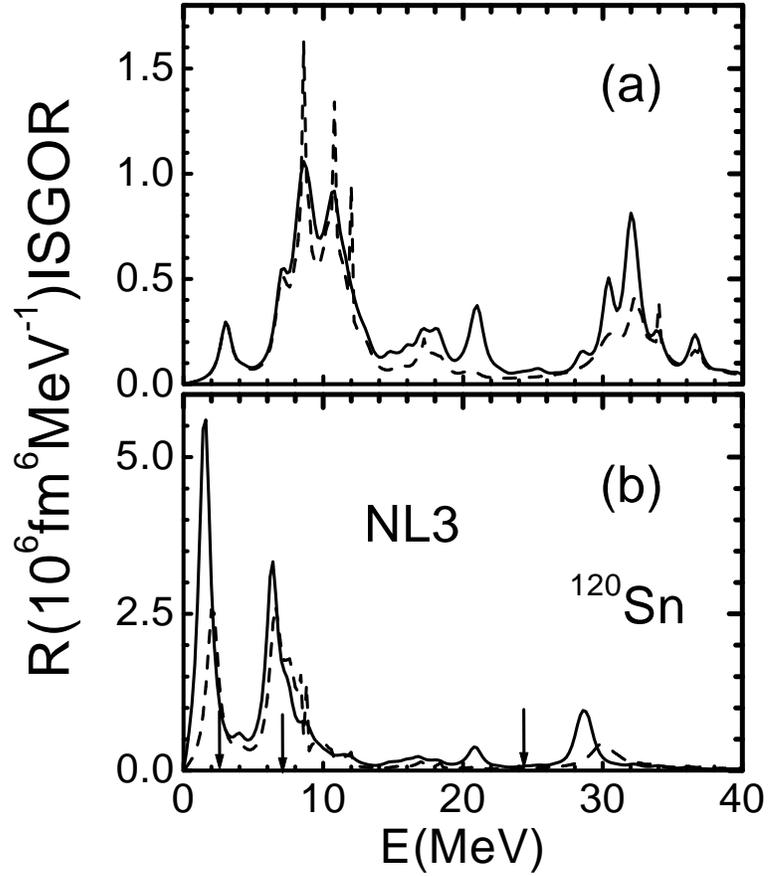}
\caption{ Hartree (a) and RRPA (b) strength as a function of the
excitation energy. The solid curves are the calculation including
all discrete particle states in the continuum and the dashed ones
are those with only the single particle resonant-states in the
continuum. The experimental data are plotted by arrows.}
\label{fig2}
\end{figure}
\end{document}